\documentclass[twocolumn,groupedaddress,showpacs,amsfonts,amssymb,amsmath]{revtex4}
\usepackage{graphicx}

\begin{document}

\bibliographystyle{plain}

\title{Multivariate phase space reconstruction\\ by nearest neighbor
embedding with different time delays}

\author{Sara P. Garcia}
\email[Email address:]{spinto@itqb.unl.pt}
\affiliation{Biomathematics Group, Instituto de Tecnologia
Qu\'imica e Biol\'ogica, Universidade Nova de Lisboa, Rua da
Quinta Grande 6, 2780-156 Oeiras, Portugal}

\author{Jonas S. Almeida}
\affiliation{Department of Biostatistics, Bioinformatics and
Epidemiology, Medical University of South Carolina, 135 Cannon
Street, Charleston, South Carolina 29425, USA}

\date{\today}

\begin{abstract}
A recently proposed nearest neighbor based selection of time
delays for phase space reconstruction is extended to multivariate
time series, with an iterative selection of variables and time
delays. A case study of numerically generated solutions of the
$x-$ and $z$ coordinates of the Lorenz system, and an application
to heart rate and respiration data, are used for illustration.\\
\begin{center}
(\copyright 2005 The American Physical Society,
\href{http://link.aps.org/abstract/PRE/v72/e027205}{http://link.aps.org/abstract/PRE/v72/e027205})
\end{center}
\end{abstract}


\maketitle

Phase space reconstruction by time delay embedding is at the
center of most nonlinear time series analysis methods (see
\cite{1,2} for an introduction). It is a primal goal as it
ensures, under certain generic conditions, the reconstruction of a
phase space equivalent to the original one, thus allowing a
qualitative and quantitative analysis of the underlying dynamical
system. The most commonly used embedding techniques are based on
Takens embedding theorem \cite{3}, which only considers
delay-coordinate maps built from a single observable, that is, a
scalar time series \cite{3,4}. Even though the widespread use and
paramount importance of this embedding theorem, it can be
extremely difficult to reconstruct a phase space from a scalar
time series when more than a couple degrees of freedom are active
\cite{5}, which is the common scenario when analyzing biological
systems, due to the complexity of their structures and complicated
dynamics. The motivation for this study is the increasing interest
in reverse-engineering biological systems directly from time
series (see \cite{6} for example), whose typical multivariate,
finite and noisy nature renders it particularly important to
develop efficient multivariate embedding schemes \cite{7}.

Generically consider a smooth deterministic dynamical system
$s(t)=f(s(t_0))$, either in continuous or discrete time, whose
trajectories are asymptotic to a compact $d$-dimensional manifold.
By performing $k$-dimensional measurements on the system, where
$k=1, \ldots, d$, it is possible to define a function
$\mathbf{x}_{(i)}=h[s(t=i \times \delta)]$ that relates the states
of the dynamical system throughout time with a time series of
measured points, where $\mathbf{x}_{(i)} \in \mathbb{R}^{k}$,
$i=1, \ldots, n$; $n$ is the total number of sampled points, and
$\delta$ is the sampling time. Phase space reconstruction by time
delay embedding is a method of generating an $m$-dimensional
manifold, from the $(n \times k)$ available measurements, that is
equivalent to the original $d$-dimensional manifold. In the scalar
scenario, that is, for $k=1$, an $m$-dimensional embedding matrix
of delay-coordinate column vectors can be defined from the initial
time series $\mathbf{x}_{(i)}$, as $\mathbf{X}=[\mathbf{x}_{(i)},
\mathbf{x}_{(i+\tau_1)}, \mathbf{x}_{(i+\tau_2)}, \dots,
\mathbf{x}_{(i+ \tau_{(m-1)})}]$. Building such a matrix implies
estimating two parameters: the time delay $\tau$, which is the
time displacement between successive columns, and the embedding
dimension $m$, which is the dimension, or number of columns, of
the final embedding matrix. We have recently proposed \cite{8} a
nearest neighbor measure $N$ for time delay embedding, solely
based on topological and dynamical arguments documented by the
data. This measure possesses the useful feature of retaining the
inverse relationship with structure disclosure, meaning that it
first decreases with $\tau$, and then returns to higher values
when $\tau$ is too long for dynamic coupling to be retained. When
the time series is noise-free, such $\tau$ value corresponds to
the global minimum of $N$ and an upper limit to an efficient
selection of time delays, beyond which statistical independence
reflects dynamic decoupling. Furthermore, it was shown \cite{8}
that using different time delays for consecutive embedding
dimensions is more efficient than using the same $\tau$ value for
all dimensions, which has been the common approach to phase space
reconstruction by time delay embedding. Hence, the $N$ algorithm
will output a vector of different time delays $[\tau_1, \tau_2,
\ldots, \tau_{(m-1)}]$, as incorporated in the definition of
embedding matrix above. In this report we extend that nearest
neighbor embedding with different time delays method to
multivariate time series, that is, when $1 < k < d$, by selecting,
at each iteration, the combination of variable, from the initial
set of $k$ variables, and time delay that first minimizes $N$. As
before \cite{8}, the false nearest neighbors (F) algorithm
proposed by Kennel \emph{et al.} \cite{9} is used to set the final
embedding dimension. This algorithm considers the ratio of
Euclidean distances between each and every point and its nearest
neighbor, first on a $m$-dimensional and then on a
$(m+1)$-dimensional space. If the ratio is greater than a given
threshold, these points are referred to as false nearest
neighbors, that is, points that appear to be nearest neighbors not
because of the dynamics, but because the attractor is being viewed
in an embedding space too small to unfold it. When the fraction of
F as a function of the embedding dimension decreases to zero, the
underlying attractor is unfolded and $m$ can be optimally
estimated.

The nearest neighbor embedding with different time delays
algorithm for multivariate time series is detailed below. (i)
Consider an initial multivariate time series $\mathbf{x}_{[i
\times k]}$, $i=1, \ldots, n$, of $n$ measurements of $k$
different dynamical variables. For each variable and for each
$\tau$ being tested, build a candidate embedding matrix
$\mathbf{T}$:
\begin{tabbing}
\hspace{0.1in} \= \hspace{0.1in} \= \kill
\verb"for" $j$ \textsf{from} 1 \textsf{to} $k$ $\{$ \> \> \\
    \> \verb"for" $\tau$ \textsf{from} 1 \textsf{to} $\frac{1}{10} n$ $\{$ \verb"define" $\mathbf{T}=[\mathbf{x}_{(i,1:k)}, \mathbf{x}_{(i+\tau,j)}]$ $\}$ \\
$\}$ \>
\end{tabbing}
(the upper limit for $\tau$ is chosen arbitrarily). (ii) For each
$(k+1)$-dimensional point, that is, for each row in matrix
$\mathbf{T}$, estimate its $(k+1)$-dimensional nearest neighbor.
Calculate the Euclidean distance between them $d_{E1}$. (iii)
Consider both points one sampling unit ahead, still in
$(k+1)$-dimensions, and calculate the new Euclidean distance
between them $d_{E2}$. (iv) Estimate the ratio $d_{E2}/d_{E1}$ and
save the number of distance ratios larger than 10. That fraction
is what is referred to as $N$ in the $\tau$ selecting profiles
ahead. The threshold value, though heuristically set, is justified
by numerical studies \cite{9} and has low parametric sensitivity.
(v) From the profiles of $N$ vs $\tau$ for each of the $k$
variables, select the combination of variable and time delay that
first minimizes $N$. That should be the optimal variable $j_1$ and
time delay $\tau_1$ selection for this first embedding cycle (that
is, each iteration that adds another dimension to the candidate
embedding matrix). (vi) Estimate the percentage of F and save that
value as a function of the dimensionality of the candidate
embedding matrix $\mathbf{T}$. (vii) Build a putative embedding
matrix $\mathbf{X}=[\mathbf{x}_{(i,1:k)},
\mathbf{x}_{(i+\tau_1,j_1)}]$, where $1 \leq j_1 \leq k$. (viii)
Again, for each variable and $\tau$ being tested, build a
candidate embedding matrix, now $\mathbf{T}=[\mathbf{x}_{(i,1:k)},
\mathbf{x}_{(i+\tau_1,j_1)}, \mathbf{x}_{(i+\tau,j)}]$. (ix)
Repeat steps (ii) to (viii), considering that now points are
$(k+2)$- and more dimensional. (x) Stop this iterative procedure
when the fraction of F [step (vi)] has dropped to 0. The outcome
of running this procedure for as long as necessary to minimize F
is the final $m$-dimensional embedding matrix:
$\mathbf{X}=[\mathbf{x}_{(i,1:k)}, \mathbf{x}_{(i+\tau_1,j_1)},
\mathbf{x}_{(i+\tau_2,j_2)},\ldots,
\mathbf{x}_{[i+\tau_{(m-k)},j_{(m-k)}]}]$, where $\tau \in
\mathbb{N}$ and $1 \leq j \leq k$.

Two bivariate data sets will be used to illustrate the
multivariate extension of the nearest neighbor embedding with
different time delays. The first are the $x-$ and $z$ coordinates
[$L(X)$ and $L(Z)$, respectively] of the Lorenz system of
differential equations \cite{10} $\dot{x}=\sigma(y-x)$,
$\dot{y}=x(\rho-z)-y$, $ \dot{z}=xy-\beta z$, with parameters
$\sigma = 10, \rho = 28, \beta = 8/3$. The equations were
numerically integrated with a 4--5th order Runge-Kutta algorithm,
sampled at $\delta = 0.01$ intervals, and transients were removed.
The second data set is composed by two physiological signals, the
heart rate [$P(H)$] and respiration [$P(R)$] from a 49-year-old
man diagnosed with sleep apnea, a potentially life-threatening
disorder in which the subject stops breathing during sleep. The
data was extracted from data set B of the 1991 \emph{Santa Fe Time
Series Prediction and Analysis Competition} \cite{11}. The
variables were digitized at 250 Hz and then sampled at 0.5 second
intervals. The units of the $P(H)$ measurements are beats per
minute, while $P(R)$ is provided in uncalibrated digitization
units (see \cite{11} for more detailed information on the data set
and its pre-processing). When considering multivariate time
series, normalization is a pivotal pre-requisite to overcome scale
shifts. Accordingly, we have non-parametrically normalized each
variable separately to its empirical cumulative distribution, by
first sorting all $n$ values and then replacing them by their
$rank/n$. Each data set, as used in the subsequent analysis,
includes a total of 8000 points, part of which is plotted in Fig.
1. The section of data set B used in this report includes both a
period of apnea and a period of intermittent apnea.

\begin{figure}[h]
\includegraphics{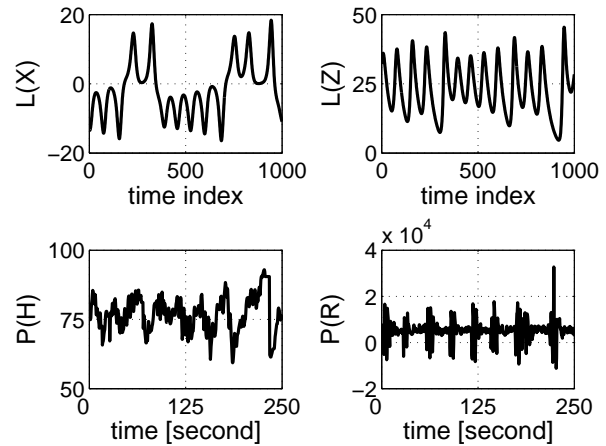}
\caption{\label{Fig. 1.} Data sets. Upper panel: the $x$ [$L(X)$,
left] and $z$ [$L(Z)$, right] coordinates of the Lorenz system.
Lower panel: heart rate [$P(H)$, left] and respiration [$P(R)$,
right] signals, in beats per minute and uncalibrated digitization
units, respectively. In the subsequent analysis, each variable is
normalized to its empirical cumulative distribution.}
\end{figure}

First, consider the case study of two coordinates of the Lorenz
system. As this is a low-dimensional system, there is no obvious
advantage in using multivariate time series to reconstruct the
phase space. Therefore, this example is used only to illustrate
the multivariate procedure for a well-described system, where the
problems of noise and nonstationarity, typically encountered in
biological data sets, are absent. The $N$ profiles for selecting
$\tau$ for the first embedding cycle are displayed in Fig. 2(a),
where the thick line indicates the profile for $L(X)$, meaning
that the candidate matrix being tested is $[L(X)_{(i)},
L(Z)_{(i)}, L(X)_{(i+\tau)}]$, and the thin line indicates the
profile for $L(Z)$, meaning that the candidate matrix being tested
is $[L(X)_{(i)}, L(Z)_{(i)}, L(Z)_{(i+\tau)}]$. The variable
$L(X)$ and the $\tau$ value of its $N$ profile first minimum
$\tau_1$ are the optimal combination that is selected from this
first embedding cycle [Fig. 2(a), circle], which implies that the
candidate embedding matrices that would be tested in a second
embedding cycle would be $[L(X)_{(i)}, L(Z)_{(i)},
L(X)_{(i+\tau_1)}, L(X)_{(i+\tau)}]$ and $[L(X)_{(i)}, L(Z)_{(i)},
L(X)_{(i+\tau_1)}, L(Z)_{(i+\tau)}]$. The rationale for using the
first minimum was discussed in \cite{8}, where it was shown to be
the most efficient choice. Displayed in Fig. 2(b) is the fraction
of F as a function of $m$ \cite{9}, from which can be concluded
that the optimal embedding dimension is $m=3$, and the final
embedding matrix is then $[L(X)_{(i)}, L(Z)_{(i)},
L(X)_{(i+\tau_1)}]$.

\begin{figure}[h]
\includegraphics{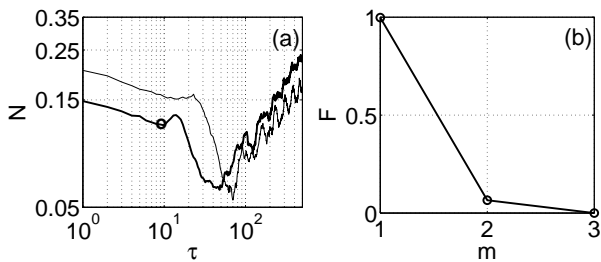}
\caption{\label{Fig. 2.} (a) First embedding cycle profiles for
$\tau$ selection from $N$. The thick line indicates the profile
for $L(X)$ and the thin line the profile for $L(Z)$. A circle
indicates the optimal combination of variable and time delay that
is selected from this first cycle. The global minima of $N$
corresponds to the onset of dynamic decoupling. (b) The fraction
of F as a function of $m$ for the embedding of the bivariate
$L(X)$ and $L(Z)$ time series. See the text for a more
comprehensive description.}
\end{figure}

\begin{figure*}[]
\includegraphics{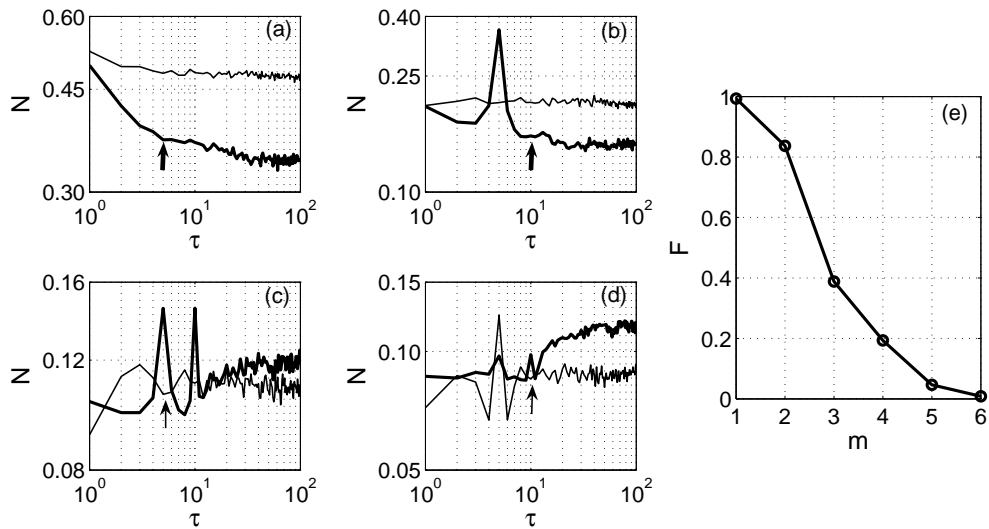}
\caption{\label{Fig. 3.} (a) First, (b) second, (c) third and (d)
fourth embedding cycle profiles for $\tau$ selection from $N$.
Thick lines indicate profiles for $P(H)$ and thin lines indicate
profiles for $P(R)$. An arrow indicates the optimal combination of
variable and time delay selected from each embedding cycle. From
the second (b) to the fourth (d) embedding cycles, a peak is
visible at previously selected $\tau$ values. (e) The fraction of
F as a function of $m$ for the embedding of the bivariate $P(H)$
and $P(R)$ time series. See the text for a more comprehensive
description.}
\end{figure*}

Consider now the biological data set. Physiological systems are
typically high-dimensional, nonstationary and contaminated by
noise. We have taken no action to correct these problems, in the
sense that we have used the data as given for the \emph{Santa Fe
Competition}. The $\tau$ selecting profiles for the first, second,
third and fourth embedding cycles are displayed in Figs. 3(a) to
3(d), respectively, where thick lines indicate the $N$ profiles
for $P(H)$ and thin lines indicate the profiles for $P(R)$. The
candidate embedding matrices are built in the same way as
described in the Lorenz case study, that is, $[P(H)_{(i)},
P(R)_{(i)}, P(H)_{(i+\tau)}]$ and $[P(H)_{(i)}, P(R)_{(i)},
P(R)_{(i+\tau)}]$ are the candidate embedding matrices being
tested for the first embedding cycle [Fig. 3(a)], and for the next
embedding cycles, the candidate matrices are built based on the
combination of variable and time delay selected in the previous
cycle [Figs. 3(a) to 3(d), arrow]. As these are noisy profiles
(see \cite{8} for a comparison with profiles for $L(X)$
contaminated with additive Gaussian white noise), the
identification of local minima becomes more difficult and is
sometimes balanced with the identification of a point at which a
change in the decaying velocity of the profile occurs. From the
second to the fourth embedding cycles [Figs. 3(b) to 3(d)], a peak
is visible at previously selected $\tau$ values, a feature also
present in the noisy profiles reported in \cite{8}, which
indicates that selecting the same $\tau$ value would not only be a
suboptimal choice, as it would indeed be the worst possible
choice. Displayed in Fig. 3(e) is the fraction of F as a function
of $m$ \cite{9}, from which can be concluded that the optimal
embedding dimension is $m=6$, and the final embedding matrix is
then $[P(H)_{(i)}, P(R)_{(i)},
P(H)_{(i+\tau_1)}, P(H)_{(i+\tau_2)}, P(R)_{(i+\tau_3)},\\
P(R)_{(i+\tau_4)}]$. The embedding of variables with different
oscillatory frequencies, such as the heart rate $P(H)$ and
respiration $P(R)$, will initially be biased towards the variable
with the higher frequency. This is clearly visible in Fig. 3(a),
with $P(H)$ presenting a substantially lower $N$ profile than that
of $P(R)$. However, this initial selection of $P(H)$ is
increasingly less advantageous, until the alternative variable
$P(R)$ is favored for the efficiency of the embedding.

The multivariate phase space reconstruction scheme could have been
conceived in different ways, from embedding each variable
separately and then adding them together, to the proposed
iterative selection from an initial set of variables. The latter
is more efficient, in the sense that the final embedding dimension
is smaller then when variables are embedded separately. Such
advantage is particularly relevant for massively multivariate
systems, as for example proteomics or transcriptomic time series,
which include hundreds or even thousands of variables \cite{12}.
Many of these variables will likely have a very strong correlation
among themselves. In that case, the most efficient phase space
reconstruction does not necessarily start with the concatenation
of all variables without delay, as the approach suggested in this
report. Instead, it should start with a single variable, to which
additional variables, first without any delay, are then added.
This small variation to the proposed algorithm addresses the issue
of sufficient representation in multivariate systems with tight
correlation between parameters. It is interesting to note that the
proposed implementation in fact treats each variable as a
surrogate for a delayed representation of the other variables.
This is also particularly well suited for the representation of
dynamic behavior documented by molecular biology time series for a
very pragmatical reason -– they tend to be very short, in the
sense that the number of time points is typically many fold
smaller then the number of variables.

The reverse engineering of biological processes from the time
series they generate is often approached by parametrization of an
explicit mathematical formulation \cite{6}. It can be argued that
this approach is hampered by the lack of exploratory tools that
analyze the dynamic behavior directly to assist in selecting the
most explanatory independent variables. Furthermore, the
characteristic heterogeneity in oscillatory frequencies, large
number of variables and short duration of the time series creates
a particular challenge for approaching this exploratory analysis
through the characterization of a reconstructed attractor.
Accordingly, this report describes an attempt to use the criteria
of efficient embedding to achieve that goal.

\begin{acknowledgments} Supported by Grants No.
SFRH/BD/1165/2000 and No. POCTI/1999/BSE/34794 from Funda\c c\~ao
para a Ci\^encia e a Tecnologia, Portugal, and by the National
Heart, Lung and Blood Institute (NIH) Proteomics Initiative
through Contract No. N01-HV-28181 (D. Knapp, PI).
\end{acknowledgments}

\bibliography{GarciaAlmeidaNembeddingmultivar}

\begin{thebibliography}{10}

\bibitem{2}
H.D.I. Abarbanel.
\newblock {\em Analysis of Observed Chaotic Data}.
\newblock Springer, New York, 1996.

\bibitem{12}
M.Y. Galperin.
\newblock {\em Nucl.\ Acids Res.}, 33:D5--D24, 2005.

\bibitem{8}
S.P. Garcia and J.S. Almeida.
\newblock Nearest neighbor embedding with different time delays.
\newblock {\em Phys.\ Rev.\ E}, 71:037204, 2005.

\bibitem{7}
R.~Hegger, L.~Jaeger, and H.~Kantz.
\newblock Reconstruction of the dynamics of noisy multivariate time series.
\newblock Internal report, Max-Planck-Institut $\textrm{f\"ur}$ Physik
  komplexer Systeme, 1997.

\bibitem{5}
H.~Kantz, J.~Kurths, and G.~Mayer-Kress (Eds.).
\newblock {\em Nonlinear Analysis of Physiological Data}.
\newblock Springer, Heidelberg, 1998.

\bibitem{1}
H.~Kantz and T.~Schreiber.
\newblock {\em Nonlinear Time Series Analysis}.
\newblock Cambridge University Press, Cambridge, UK, 1997.

\bibitem{9}
M.B. Kennel, R.~Brown, and H.D.I. Abarbanel.
\newblock {\em Phys.\ Rev.\ A}, 45:3403--3411, 1992.

\bibitem{10}
E.~Lorenz.
\newblock {\em J.\ Atmos.\ Sci.}, 20:130--141, 1963.

\bibitem{11}
D.R. Rigney, A.L. Goldberger, W.C. Ocasio, Y.~Ichimaru, G.B. Moody, and R.G.
  Mark.
\newblock {\em Time Series Prediction: Forecasting the Future and Understanding
  the Past}, volume Proc. Vol. XV of {\em SFI Studies in the Sciences of
  Complexity}, chapter Multi-channel physiological data: description and
  analysis (Data Set B), pages 105--129.
\newblock Perseus Books, Reading, MA, 1st edition edition, 1994.

\bibitem{4}
T.~Sauer, J.A. Yorke, and M.~Casdagli.
\newblock {\em J.\ Stat.\ Phys.}, 65:579--616, 1991.

\bibitem{3}
F.~Takens.
\newblock {\em Dynamical Systems and Turbulence}, volume 898 of {\em Lecture
  Notes in Mathematics}, pages 366--381.
\newblock Springer-Verlag, Berlin, 1981.

\bibitem{6}
E.~Voit and J.~Almeida.
\newblock Decoupling dynamical systems for pathway identification from
  metabolic profiles.
\newblock {\em Bioinformatics}, 20(11):1670--1681, 2004.

\end{thebibliography}

\end{document}